\pdfoutput=1
\documentclass[english,journal]{IEEEtran}
\usepackage[T1]{fontenc}
\usepackage{ae}
\usepackage{aecompl}
\usepackage[latin9]{inputenc}
\usepackage{textcomp}
\usepackage{amsmath}
\usepackage{amssymb}
\usepackage{stackrel}
\usepackage{graphicx}

\makeatletter

\providecommand{\tabularnewline}{\\}

\makeatother

\usepackage{babel}
\begin{document}

\title{Null Space Based Preemptive Scheduling For Joint URLLC and eMBB Traffic
in 5G Networks}

\author{\IEEEauthorblockN{Ali A. Esswie$^{1,2}$,\textit{ Member, IEEE}, and\textit{ }Klaus
I. Pedersen\textit{$^{1,2}$, Senior Member, IEEE}\\
$^{1}$Nokia Bell-Labs, Aalborg, Denmark\\
$^{2}$Department of Electronic Systems, Aalborg University, Denmark}}

\maketitle
$\pagenumbering{gobble}$
\begin{abstract}
In this paper, we propose a null-space-based preemptive scheduling
framework for cross-objective optimization to always guarantee robust
URLLC performance, while extracting the maximum possible eMBB capacity.
 The proposed scheduler perpetually grants incoming URLLC traffic
a higher priority for instant scheduling. In case that radio resources
are not immediately schedulable, proposed scheduler forcibly enforces
an artificial spatial user separation, for the URLLC traffic to get
instantly scheduled over shared resources with ongoing eMBB transmissions.
A pre-defined reference spatial subspace is constructed for which
scheduler instantly picks the active eMBB user whose precoder is the
closest possible. Then, it projects the eMBB precoder \textit{on-the-go}
onto the reference subspace, in order for its paired URLLC user to
orient its decoder matrix into one possible null space of the reference
subspace. Hence, a robust decoding ability is always preserved at
the URLLC user, while cross-maximizing the ergodic capacity. Compared
to the state-of-the-art proposals from industry and academia, proposed
scheduler shows extreme URLLC latency robustness with significantly
improved overall spectral efficiency. Analytical analysis and extensive
system level simulations are presented to support paper conclusions. 

\end{abstract}

\section{Introduction}

Emerging fifth generation (5G) systems are envisioned to support two
major service classes: ultra-reliable low-latency communications (URLLC)
and enhanced mobile broadband (eMBB) {[}1{]}. URLLC refer to the future
services that demand extremely reliable and low latency data communication,
i.e., one-way latency up to 1 ms with $10^{-5}$ outage probability
{[}2{]}. That is, the quality of service (QoS) of the URLLC-type applications
is infringed if more than one packet out of $10^{5}$ packets are
not successfully decoded within the 1 ms deadline. This URLLC QoS
is immensely different from that of the current long term evolution
(LTE) technology {[}3{]}, where the overall spectral efficiency (SE)
is the prime objective. 

To satisfy such stringent latency requirements, the system should
be always engineered so that blocking a URLLC packet is a very rare
event. Therefore, URLLC services must satisfy their individual outage
capacity, instead of the ergodic capacity. That is, by setting an
ultra-tight target block error rate (BLER) to always ensure a sufficient
URLLC decoding ability. This way, it leads to a significant loss of
the network SE due to the fundamental tradeoff between reliability,
latency and the achievable SE {[}4{]}. 

In the recent literature, diverse 5G scheduling contributions have
been introduced. User-centric scheduling with variable transmission
time intervals (TTIs) {[}5{]} is essential to minimize the URLLC frame
alignment and queuing delays. Furthermore, URLLC spatial diversity
techniques are vital to preserve a sufficient URLLC signal-to-interference-noise-ratio
(SINR) at all times. For example, the work in {[}6{]} demonstrates
that a 4 \texttimes{} 4 multi-input multi-output (MIMO) microscopic
diversity and two orders of macroscopic diversity are required to
reach the URLLC outage SINR level. A recent study {[}7{]} further
extends the usage of the spatial diversity for URLLC by flexibly allocating
coded segments of the URLLC payload message to different interfaces.
Thus, a better latency-reliability tradeoff can be achieved by reducing
the original payload transmission time. Additionally, URLLC punctured
scheduling (PS) {[}8{]} is a state-of-the-art scheme to further minimize
the queuing delay of the URLLC traffic, where sporadic URLLC traffic
is instantly scheduled by overwriting part of the radio resources,
monopolized by ongoing eMBB transmissions.

However, the majority of the URLLC scheduling studies consider a monotonic
optimization structure of the URLLC outage capacity. Therefore, URLLC
requirements can be proportionally satisfied only with the size of
the URLLC granted resources or received SINR levels. However, when
joint eMBB and URLLC traffic coexists on the same radio spectrum,
this approach fails to reach a proper system ergodic capacity.

In this work, a null-space-based preemptive scheduling (NSBPS) for
joint eMBB and URLLC traffic is proposed. Proposed scheduler seeks
to dynamically fulfill a jointly constrained objective, for which
the URLLC QoS is guaranteed, while achieving the best possible eMBB
capacity. If the available radio resources are not sufficient to accommodate
the URLLC payload, NSBPS forcibly fits the URLLC traffic within an
ongoing eMBB transmission in an instant, controlled, semi-transparent
and biased multi-user MIMO (MU-MIMO) transmission. The proposed NSBPS
instantly selects an active eMBB user whose transmission is most aligned
within an arbitrary reference subspace. It spatially projects the
selected eMBB transmission onto the reference subspace for which its
paired URLLC user de-orients its decoding matrix into one possible
null-space. Accordingly, a robust SINR level is preserved at the URLLC
user side. Compared to the state-of-the-art studies, proposed NSBPS
shows extreme robustness of the URLLC QoS with significantly improved
ergodic capacity. 

Due to the complexity of the 5G new radio (NR) system model {[}1-3{]}
and addressed problems therein, the performance of the proposed scheduler
is validated by extensive system simulations (SLS), and supported
by analytical analysis of the major performance indicators. Those
simulations are based on widely accepted models and calibrated against
the 5G NR specifications to ensure highly reliable statistical results. 

\textit{Notations:} $(\text{\ensuremath{\mathcal{X}}})^{\textnormal{T}}$
, $(\text{\ensuremath{\mathcal{X}}})^{\textnormal{H}}$ and $(\text{\ensuremath{\mathcal{X}}})^{\textnormal{-1}}$
stand for the transpose, Hermitian, and inverse operations of $\text{\ensuremath{\mathcal{X}}}$,
$\text{\ensuremath{\mathcal{X}}}\,\cdotp\,\mathcal{Y}$ is the dot
product of $\text{\ensuremath{\mathcal{X}}}$ and $\text{\ensuremath{\mathcal{Y}}}$,
while $\overline{\mathcal{X}}$ and $\left\Vert \mathcal{X}\right\Vert $
represent the mean and 2-norm of $\text{\ensuremath{\mathcal{X}}}.$
$\ensuremath{\mathcal{X}\sim\mathbb{C\ensuremath{\mathbb{N}}}}(0,\sigma^{2})$
indicates a complex Gaussian random variable with zero mean and variance
$\sigma^{2}$, $\mathcal{X}_{\kappa},\kappa\text{\ensuremath{\in}}\{\textnormal{llc},\textnormal{mbb}\}$
denotes the type of user $\text{\ensuremath{\mathcal{X}}}$, $\mathbb{E}\left\{ \mathcal{X}\right\} $
and $\mathbf{card}(\text{\ensuremath{\mathcal{X}}})$ are the statistical
expectation and cardinality of $\text{\ensuremath{\mathcal{X}}}$.

The paper is organized as follows. Section II presents the system
and signal models, respectively. Section III states the problem formulation
and detailed description of the NSBPS scheduler. Extensive system
level simulation results are introduced in Section IV, and paper is
concluded in Section V. 

\section{System Model}

We consider a 5G-NR downlink (DL) MU-MIMO system where there are $C$
cells, each equipped with $N_{t}$ transmit antennas, and $K$ uniformly
distributed user equipment's (UEs) per cell, each with $M_{r}$ receive
antennas. Users are multiplexed by the orthogonal frequency division
multiple access (OFDMA). Two types of DL traffic are under assessment
as: (a) URLLC bursty FTP3 traffic model with a finite $\textnormal{B}-$byte
payload and Poisson arrival process $\lambda$, and (b) eMBB full
buffer traffic with infinite payload size. The total number of UEs
per cell is: $K_{\textnormal{mbb}}+K_{\textnormal{llc}}=K$, where
$K_{\textnormal{mbb}}$ and $K_{\textnormal{llc}}$ are the average
numbers of eMBB and URLLC UEs per cell, respectively. 

The agile 5G-NR frame structure is adopted {[}5{]}, where the URLLC
and eMBB UEs are scheduled with variable TTI periodicity. As depicted
in Fig. 1, eMBB traffic is scheduled with a long TTI of 14-OFDM symbols
for SE maximization while URLLC traffic with a shorter TTI of 2-OFDM
symbols due to its latency budget. In the frequency domain, the smallest
scheduling unit is the physical resource block (PRB), which is 12
sub-carriers and with 15 kHz sub-carrier spacing. 

A maximal subset of MU co-scheduled URLLC-eMBB user pairs $\text{\ensuremath{\mathtt{G}}}_{c}\text{\,\ensuremath{\in}\ \ensuremath{\mathcal{K}}}_{c}$
is allowed over an arbitrary PRB in the $c^{th}$ cell, where $G_{c}=\mathbf{card}(\text{\ensuremath{\mathtt{G}}}_{c}),\,G_{c}\leq N_{t}$
is the number of co-scheduled UEs and $\text{\ensuremath{\mathcal{K}}}_{c}$
is the set of all active UEs in the $c^{th}$ cell. Since $N_{t}\leq KM_{r}$,
user selection on top of equal power allocation is assumed for MU
pairing. The received DL signal at the $k^{th}$ user from the $c^{th}$
cell can be modeled as

\[
\boldsymbol{\textnormal{y}}_{k,c}^{\kappa}=\boldsymbol{\textnormal{\textbf{H}}}_{\kappa,c}^{\kappa}\boldsymbol{\textnormal{\textbf{v}}}_{k,c}^{\kappa}s_{\kappa,c}^{\kappa}+\sum_{g\in\mathtt{G}_{c},g\neq k}\mathit{\boldsymbol{\textnormal{\textbf{H}}}}_{k,c}\boldsymbol{\textnormal{\textbf{v}}}_{g,c}s_{g,c}
\]

\begin{equation}
+\sum_{j=1,j\neq c}^{C}\,\,\sum_{g\in\mathtt{G}_{j}}\boldsymbol{\textnormal{\textbf{H}}}_{g,j}\boldsymbol{\textnormal{\textbf{v}}}_{g,j}s_{g,j}+\boldsymbol{\textnormal{\textbf{n}}}_{k,c}^{\kappa},
\end{equation}
where $\boldsymbol{\textnormal{\textbf{H}}}_{k,c}^{\kappa}\in\text{\ensuremath{\mathcal{C}}}^{M_{r}\times N_{t}},\forall k\in\{1,\ldots,K\},\forall c\in\{1,\ldots,C\}$
is the wireless channel observed at the $k^{th}$ user from the $c^{th}$
cell, $\boldsymbol{\textnormal{\textbf{v}}}_{k,c}^{\kappa}\in\text{\ensuremath{\mathcal{C}}}^{N_{t}\times1}$
is the zero-forcing precoding vector, assuming a single layer transmission
per user, where it is given as: $\boldsymbol{\textnormal{\textbf{v}}}_{k,c}^{\kappa}=\left(\boldsymbol{\textnormal{\textbf{H}}}_{\kappa,c}^{\kappa}\right)^{\textnormal{H}}\left(\boldsymbol{\textnormal{\textbf{H}}}_{k,c}^{\kappa}\left(\boldsymbol{\textnormal{\textbf{H}}}_{\kappa,c}^{\kappa}\right)^{\textnormal{H}}\right)^{-1}.$
$s_{k,c}^{\kappa}$ and $\boldsymbol{\textnormal{\textbf{n}}}_{\kappa,c}^{\kappa}$
denote the transmitted symbol and the additive white Gaussian noise
at the $k^{th}$ user, respectively. The first and second summation
terms represent the intra-cell inter-user and inter-cell interference,
generated from either the URLLC or eMBB traffic. In this work, the
3GPP 3D spatial channel model {[}9{]} is adopted, where the DL channel
coefficient observed by the $m^{th}$ receive antenna from the $n^{th}$
transmit antenna is composed from $Q$ spatial clusters, each with
$Z$ rays as
\begin{figure}
\begin{centering}
\includegraphics[scale=0.6]{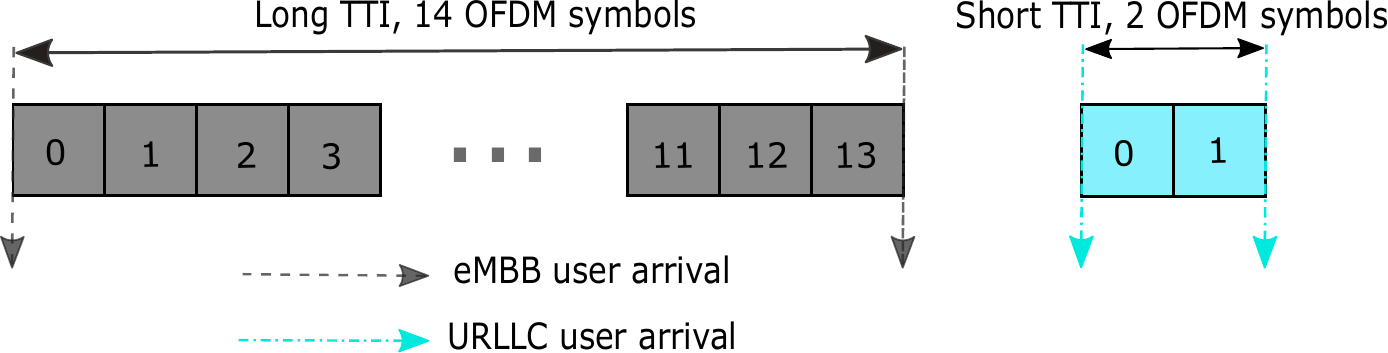}
\par\end{centering}
\centering{}{\small{}Fig. 1. Flexible TTI scheduling in 5G NR.}{\small \par}
\end{figure}

\begin{equation}
h_{(m,n)_{k}}^{\kappa}=\frac{1}{\sqrt{Q}}\sum_{q=0}^{Q-1}\sqrt{\delta_{k}}\text{\,\ensuremath{\mathcal{G}}}_{q,k}\,r_{(m,n,q)_{k}},
\end{equation}
where $\delta_{k}=\ell\epsilon_{k}^{\text{\ensuremath{\varrho}}}\mu_{k}$,
$\ell$ and $\mu_{k}$ are the propagation and shadow fading coefficients,
respectively, and $\epsilon_{k}^{\varrho}$ is the distance, with
$\text{\ensuremath{\varrho}}$ as the pathloss factor, and $\text{\ensuremath{\mathcal{G}}}_{q,k}\sim\text{\ensuremath{\mathbb{C\ensuremath{\mathbb{N}}}}(0,1). }$The
steering factor $r_{(m,n,q)_{k}}$ is given by 

$r_{(m,n,q)_{k}}=$

\begin{equation}
\sqrt{\frac{\xi\psi}{Z}}\sum_{z=0}^{Z-1}\left(\begin{array}{c}
\sqrt{\text{\ensuremath{\mathfrak{D}}}_{\textnormal{BS}}^{m,n,q,z}(\theta_{\textnormal{AoD}},\varphi_{\textnormal{EoD}})}\,e^{j\textnormal{(\text{\ensuremath{\eta}}\textit{d}\ensuremath{\overline{f}}+\ensuremath{\Phi_{m,n,q,z}})}}\\
\times\sqrt{\text{\ensuremath{\mathfrak{D}}}_{\textnormal{UE}}^{m,n,q,z}(\theta_{\textnormal{AoA}},\varphi_{\textnormal{EoA}})}\,e^{j(\text{\ensuremath{\eta}}d\sin(\theta_{\textnormal{m,n,q,z,AoA}}))}\\
\times e^{j\text{\ensuremath{\eta}}||s||\cos(\varphi_{\textnormal{m,n,q,z,EoA}})\cos(\theta_{\textnormal{m,n,q,z,AoA}}-\theta_{s})t}
\end{array}\right),
\end{equation}
where $\xi$ and $\psi$ are the power and large-scale coefficients,
$\text{\ensuremath{\mathfrak{D}}}_{\textnormal{BS}}$ and $\text{\ensuremath{\mathfrak{D}}}_{\textnormal{UE}}$
are the antenna patterns at the BS and UE, respectively, $\text{\ensuremath{\eta}}$
is the wave number, $\theta$ denotes the horizontal angle of arrival
$\theta_{\textnormal{AoA}}$ and departure $\theta_{\textnormal{AoD}}$,
while $\varphi$ denotes the elevation angle of arrival $\varphi_{\textnormal{EoA}}$
and departure $\varphi_{\textnormal{EoD}}$, respectively. $s$ is
the speed of the $k^{th}$ user, $\overline{f}=f_{x}\cos\theta_{\textnormal{AoD}}\,\cos\varphi_{\textnormal{EoD}}$
is the displacement vector of the uniform linear transmit array.

The received signal at the $k^{th}$ user is decoded by applying the
antenna combining as: $\left(\boldsymbol{\textnormal{y}}_{k,c}^{\kappa}\right)^{*}=\left(\boldsymbol{\textnormal{\textbf{u}}}_{k,c}^{\kappa}\right)^{\textnormal{H}}\boldsymbol{\textnormal{y}}_{k,c}^{\kappa},$
where $\boldsymbol{\textnormal{\textbf{u}}}_{k,c}^{\kappa}$ is designed
by the linear minimum mean square error interference rejection combining
(LMMSE-IRC) receiver {[}10{]}. The received SINR level at the $k^{th}$
user is then calculated as

{\small{}
\begin{equation}
\varUpsilon_{k,c}^{\kappa}=\frac{p_{k}^{c}\left\Vert \boldsymbol{\textnormal{\textbf{H}}}_{k,c}^{\kappa}\boldsymbol{\textnormal{\textbf{v}}}_{k,c}^{\kappa}\right\Vert ^{2}}{1+\underset{g\in\text{\ensuremath{\mathtt{G}}}_{c},g\neq k}{\sum}p_{g}^{c}\left\Vert \boldsymbol{\textnormal{\textbf{H}}}_{k,c}^{\kappa}\boldsymbol{\textnormal{\textbf{v}}}_{g,c}^{\kappa}\right\Vert ^{2}+\underset{j\in C,j\neq c\,}{\sum}\underset{g\in\text{\ensuremath{\mathtt{G}}}_{j}}{\sum}p_{g}^{j}\left\Vert \boldsymbol{\textnormal{\textbf{H}}}_{g,j}^{\kappa}\boldsymbol{\textnormal{\textbf{v}}}_{g,j}^{\kappa}\right\Vert ^{2}},
\end{equation}
}where $p_{k}^{c}$ is the transmit power intended for the $k^{th}$
user. Then, the $k^{th}$ user received rate on a given PRB is given
by\\
\begin{equation}
r_{_{k,rb}}^{\kappa}=\log_{2}\left(1+\frac{1}{G_{k,c}}\varUpsilon_{k,c}^{\kappa}\right).
\end{equation}
Accordingly, the user SINR levels across different $\text{\ensuremath{\mathcal{N}}}$
sub-carriers are mapped into a single effective SINR using the effective
exponential SNR mapping {[}11{]} as

\begin{equation}
\left(\varUpsilon_{k,c}^{\kappa}\right)^{\textnormal{eff.}}=-\partial\ln\left(\frac{1}{\text{\ensuremath{\mathcal{N}}}}\stackrel[i=1]{\text{\ensuremath{\mathcal{N}}}}{\sum}\,e^{-\frac{\left(\varUpsilon_{k,c}^{\kappa}\right)^{i}}{\partial}}\right),
\end{equation}
with $\text{\ensuremath{\partial}}$ as a calibration parameter. 

\section{Proposed NSBPS Scheduler}

\subsection{Problem Formulation}

Under a 5G-NR system, there are user-centric, instead of network-centric,
QoS utility functions. These are highly coupled and need to be reliably
fulfilled, e.g., eMBB rate maximization and URLLC latency minimization
as

\begin{equation}
\text{\ensuremath{\forall}}k_{\textnormal{mbb}}\in\text{\ensuremath{\mathcal{K}}}_{\textnormal{mbb}}:\arg\underset{\text{\ensuremath{\mathcal{K}}}_{\textnormal{mbb}}}{\max}\sum_{k_{\textnormal{mbb}}=1}^{K_{\textnormal{mbb}}}\sum_{rb\in\Xi_{k}^{\textnormal{mbb}}}\beta_{k_{\textnormal{mbb}}}r_{_{k,rb}}^{\textnormal{mbb}},
\end{equation}

{\small{}
\begin{equation}
\forall k_{\textnormal{llc}}\in\text{\ensuremath{\mathcal{K}}}_{\textnormal{llc}}:\arg\underset{\text{\ensuremath{\mathcal{K}}}_{\textnormal{llc}}}{\min}\left(\varPsi\right),
\end{equation}
}{\small \par}

\[
\textnormal{s.t.}\left\Vert \boldsymbol{\textnormal{\textbf{v}}}_{k}^{\kappa}\sqrt{\textnormal{P}}\right\Vert ^{2},\,\varPsi\leq1\,\textnormal{ms},
\]
where $\text{\ensuremath{\mathcal{K}}}_{\textnormal{mbb}}$ and $\text{\ensuremath{\mathcal{K}}}_{\textnormal{llc}}$
represent the active sets of eMBB and URLLC users, respectively, $\Xi_{k}^{\textnormal{mbb}}$
and $\beta_{k_{\textnormal{mbb}}}$ denote the granted set of PRBs
and a priority factor of the $k^{th}$ eMBB user. $\varPsi$ is the
URLLC target one-way latency, which is expressed as

\begin{equation}
\varPsi=\Lambda_{\textnormal{q}}+\Lambda_{\textnormal{bsp}}+\Lambda_{\textnormal{fa}}+\Lambda_{\textnormal{tx}}+\Lambda_{\textnormal{uep}},
\end{equation}
where $\Lambda_{\textnormal{q}},\Lambda_{\textnormal{bsp}},\Lambda_{\textnormal{fa}},\Lambda_{\textnormal{tx}},\Lambda_{\textnormal{uep}}$
are the queuing, BS processing, frame alignment, transmission, and
UE processing delays, respectively. $\Lambda_{\textnormal{fa}}$ is
upper-bounded by the short TTI interval while $\Lambda_{\textnormal{bsp}}$
and $\Lambda_{\textnormal{uep}}$ are bounded by 3-OFDM symbol duration
{[}12{]}, due to the enhanced processing capabilities with the 5G-NR.
Hence, $\Lambda_{\textnormal{tx}}$ and $\Lambda_{\textnormal{q}}$
are the major impediment against achieving the hard URLLC latency
budget. $\Lambda_{\textnormal{tx}}$ depends on the outage SINR level
as given by

\begin{equation}
\Lambda_{\textnormal{tx}}=\frac{\textnormal{B}}{\left(\Xi_{k}^{\textnormal{llc}}\log_{2}\left(1+\frac{\varUpsilon_{k}^{\textnormal{llc}}}{\digamma}\right)\right)},
\end{equation}
where $\digamma$ is the outage gap between the expected and actual
received SINR levels. The URLLC queuing delay $\Lambda_{\textnormal{q}}$
can be modeled by the $\text{\ensuremath{\mathcal{A}}/\ensuremath{\mathcal{A}}/\ensuremath{\mathtt{a}}/\ensuremath{\phi}}$
queuing model {[}13{]}, where the first $\text{\ensuremath{\mathcal{A}}}$
denotes a Poisson packet arrival, second $\text{\ensuremath{\mathcal{A}}}$
means exponential service times out of the queue, notation $\text{\ensuremath{\mathtt{a}}}$
represents the maximum number of the URLLC simultaneous transmissions,
and notation $\text{\ensuremath{\phi}}$ implies that an arriving
URLLC packet will be dropped if there are $\text{\ensuremath{\phi}}$
outstanding packets, worth of more than $1$ ms in the queue. Thus,
the probability of the URLLC reliability loss, i.e., $\Lambda_{\textnormal{q}}\geq1$
ms, is given as

\begin{equation}
\rho_{_{rl}}=\left(\rho_{_{0}}\frac{\mathtt{a}^{\mathtt{a}}}{\mathtt{a}!}\right)\rho^{\phi},
\end{equation}
where $\rho_{_{0}}$ is the probability of the queue being empty,
and $\rho=\left(\frac{\lambda}{\mathtt{a}\mathcal{O}}\right)$, with
$\frac{1}{\mathcal{O}}$ as the mean service time. Thus, to achieve
the critical URLLC latency, the transmission and queuing delays should
be always minimized to provide further allowance for the re-transmission
delay. This can be achieved by guaranteeing a sufficient outage SINR
level or allocating excessive PRBs to URLLC traffic in order to further
minimize $\rho_{_{rl}}$. In both cases, the eMBB utility function
in (7) will be ill-optimized, leading to a severe degradation of the
network SE. 

\subsection{Description of The Proposed NSBPS Scheduler}

The proposed NSBPS scheduler seeks to simultaneously cross-optimize
the joint objectives of the eMBB and URLLC traffic. Thus, the critical
URLLC latency deadline is satisfied regardless of the system loading
while reaching the best achievable eMBB performance. In the following
sub-sections, we describe the proposed NSBPS scheduler in-detail.

\textbf{\textit{At the BS side:}}

At an arbitrary TTI instance, if there are no newly incoming URLLC
packets, NSBPS allocates single-user (SU) dedicated resources to the
new/buffered eMBB traffic based on the standard proportional fair
(PF) metric as

\begin{equation}
\Theta_{\textnormal{PF}}=\frac{r_{_{k,rb}}^{\textnormal{mbb}}}{\overline{r}_{k,rb}^{\textnormal{mbb}}},
\end{equation}

\begin{equation}
k_{\textnormal{mbb}}^{\circ}=\arg\underset{\text{\ensuremath{\mathcal{K}}}_{\textnormal{mbb}}}{\max}\,\,\,\,\Theta_{\textnormal{PF}},
\end{equation}
where $\overline{r}_{k,rb}^{\textnormal{mbb}}$ is the average delivered
data rate of the $k^{th}$ user. If sporadic DL URLLC packets arrive
at the BS while sufficient radio resources are instantly available,
the NSBPS scheduler immediately overpowers the eMBB traffic SU priority
and assigns SU resources to incoming URLLC traffic based on the weighted
PF (WPF) criteria instead as: $\Theta_{\textnormal{WPF}}=\frac{r_{_{k,rb}}^{\kappa}}{\overline{r}_{k,rb}^{\kappa}}\beta_{k_{\kappa}},$
with $\beta_{k_{\textnormal{llc}}}\gg\beta_{k_{\textnormal{mbb}}}$
for immediate URLLC SU scheduling. 

However, under a large offered load, which is envisioned with the
5G-NR, schedulable resources may not be instantly available for critical
URLLC traffic and accordingly, significant queuing delays are foreseen.
In such case, NSBPS scheduler first attempts to fit the URLLC packets
within an active eMBB traffic in a normal and non-biased MU transmission,
based on a conservative $\gamma-$orthogonality threshold, where $\text{\ensuremath{\gamma}}\rightarrow\left[0,1\right]$.
Thus, the incoming URLLC traffic can only be paired with an active
eMBB user if they satisfy: 

\begin{equation}
1-\left|\left(\boldsymbol{\textnormal{\textbf{v}}}_{b}^{\textnormal{mbb}}\right)^{\textnormal{H}}\boldsymbol{\textnormal{\textbf{v}}}_{e}^{\textnormal{llc}}\right|^{2}\geq\gamma,\,b\neq e,
\end{equation}
with $\forall b\in\{1,\ldots,K_{\textnormal{mbb}}\},\forall e\in\{1,\ldots,K_{\textnormal{llc}}\}.$
The conservative orthogonality threshold is enforced to safeguard
the URLLC traffic from potential inter-user interference. However,
if the spatial degrees of freedom (SDoFs) are limited within a TTI,
i.e., system is incapable to jointly process several signals between
different transceivers on the same resources, and such orthogonality
can not instantly offered, NSBPS scheduler immediately alters the
system optimization objective into a region that satisfies the URLLC
outage requirements, while imposing minimal loss to the eMBB performance.
Thus, the scheduler enforces an instant, biased and controlled MU
transmission between URLLC-eMBB user pair. The URLLC outage is then
guaranteed by satisfying the following conditions, 

\begin{equation}
\textnormal{\textbf{rank}}\left(\left(\boldsymbol{\textnormal{\textbf{u}}}_{k}^{\textnormal{llc}}\right)^{\textnormal{H}}\,\boldsymbol{\textnormal{\textbf{H}}}_{k}^{\textnormal{llc}}\,\boldsymbol{\textnormal{\textbf{v}}}_{k}^{\textnormal{llc}}\right)\sim\textnormal{full},
\end{equation}

\begin{equation}
\textnormal{\textbf{rank}}\left(\left(\boldsymbol{\textnormal{\textbf{u}}}_{k}^{\textnormal{llc}}\right)^{\textnormal{H}}\,\boldsymbol{\textnormal{\textbf{H}}}_{k}^{\textnormal{llc}}\,\left(\boldsymbol{\textnormal{\textbf{v}}}_{k^{\diamond}}^{\textnormal{mbb}}\right)^{\prime}\right)\sim\textnormal{0},
\end{equation}
where $\left(\boldsymbol{\textnormal{\textbf{v}}}_{k^{\diamond}}^{\textnormal{mbb}}\right)^{\prime}$
denotes the updated precoder of the co-scheduled eMBB user with the
incoming URLLC user. Thus, an arbitrary discrete Fourier transform
spatial subspace $\boldsymbol{\textnormal{\textbf{v}}}_{\textnormal{ref}}(\theta)$,
pointing towards angle $\theta$, is constructed by

\begin{equation}
\boldsymbol{\textnormal{\textbf{v}}}_{\textnormal{ref}}(\theta)=\left(\frac{1}{\sqrt{N_{t}}}\right)\left[1,e^{-j2\pi\Delta\cos\theta},\ldots,e^{-j2\pi\Delta(N_{t}-1)\cos\theta}\right]^{\textnormal{T}},
\end{equation}
where {\footnotesize{}$\Delta$} is the absolute antenna spacing.
Next, the NSBPS searches for one active eMBB user whose precoder is
closest possible to the reference subspace as

\begin{equation}
k_{\textnormal{mbb}}^{\diamond}=\arg\underset{\text{\ensuremath{\mathcal{K}}}_{\textnormal{mbb}}}{\,\textnormal{min\,\,}}\textnormal{\textbf{d}}\left(\boldsymbol{\textnormal{\textbf{v}}}_{k}^{\textnormal{mbb}},\boldsymbol{\textnormal{\textbf{v}}}_{\textnormal{ref}}\right),
\end{equation}
with the Euclidean distance between $\boldsymbol{\textnormal{\textbf{v}}}_{k}^{\textnormal{mbb}}$
and $\boldsymbol{\textnormal{\textbf{v}}}_{\textnormal{ref}}$ given
by 

\begin{equation}
\textnormal{\textbf{d}}\left(\boldsymbol{\textnormal{\textbf{v}}}_{k}^{\textnormal{mbb}},\boldsymbol{\textnormal{\textbf{v}}}_{\textnormal{ref}}\right)=\frac{1}{\sqrt{2}}\left\Vert \boldsymbol{\textnormal{\textbf{v}}}_{k}^{\textnormal{mbb}}\left(\boldsymbol{\textnormal{\textbf{v}}}_{k}^{\textnormal{mbb}}\right)^{\textnormal{H}}-\boldsymbol{\textnormal{\textbf{v}}}_{\textnormal{ref}}\boldsymbol{\textnormal{\textbf{v}}}_{\textnormal{ref}}^{\textnormal{H}}\right\Vert .
\end{equation}

Then, scheduler instantly projects the precoder vector of the selected
eMBB user $\boldsymbol{\textnormal{\textbf{v}}}_{k^{\diamond}}^{\textnormal{mbb}}$
onto $\boldsymbol{\textnormal{\textbf{v}}}_{\textnormal{ref}}$ as
given by

\begin{equation}
\left(\boldsymbol{\textnormal{\textbf{v}}}_{k^{\diamond}}^{\textnormal{mbb}}\right)^{\prime}=\frac{\boldsymbol{\textnormal{\textbf{v}}}_{k^{\diamond}}^{\textnormal{mbb}}\,\cdotp\,\boldsymbol{\textnormal{\textbf{v}}}_{\textnormal{ref}}}{\left\Vert \boldsymbol{\textnormal{\textbf{v}}}_{\textnormal{ref}}\right\Vert ^{2}}\times\boldsymbol{\textnormal{\textbf{v}}}_{\textnormal{ref}},
\end{equation}
where $\left(\boldsymbol{\textnormal{\textbf{v}}}_{k^{\diamond}}^{\textnormal{mbb}}\right)^{\prime}$
is the updated eMBB user precoder. The NSBPS scheduler then instantly
schedules the incoming URLLC traffic over shared resources with the
impacted eMBB user. Since the instant precoder projection is transparent
to the victim eMBB user, it exhibits a SE projection loss. However,
eMBB loss is constrained minimum, especially under high eMBB user
load, e.g., NSBPS scheduler has a higher probability to find an eMBB
user whose precoder is originally aligned within $\boldsymbol{\textnormal{\textbf{v}}}_{\textnormal{ref}}$,
such that the instant projection process would not greatly impact
its achievable capacity. Finally, the BS acknowledges the URLLC user
by a single-bit Boolean co-scheduling indication $\alpha=1$, to be
instantly transmitted in the user-centric control channel. 

\textbf{\textit{At the URLLC user side:}}

Upon reception of $\alpha=1$, the URLLC user realizes that its granted
resources, from the scheduling grant, are shared with an active eMBB
user whose transmission is aligned within the reference subspace $\boldsymbol{\textnormal{\textbf{v}}}_{\textnormal{ref}}$.
Thus, the first-stage decoder matrix of the URLLC user is constructed
by a standard LMMSE-IRC receiver to reject the inter-cell interference
as

\begin{equation}
\left(\boldsymbol{\textnormal{\textbf{u}}}_{k}^{\textnormal{llc}}\right)^{(1)}=\left(\boldsymbol{\textnormal{\textbf{H}}}_{k}^{\textnormal{llc}}\boldsymbol{\textnormal{\textbf{v}}}_{k}^{\textnormal{llc}}\left(\,\boldsymbol{\textnormal{\textbf{H}}}_{k}^{\textnormal{llc}}\boldsymbol{\textnormal{\textbf{v}}}_{k}^{\textnormal{llc}}\right)^{\textnormal{H}}+\textnormal{\textbf{W}}\right)^{^{-1}}\,\boldsymbol{\textnormal{\textbf{H}}}_{k}^{\textnormal{llc}}\boldsymbol{\textnormal{\textbf{v}}}_{k}^{\textnormal{llc}},
\end{equation}
where the interference covariance matrix is given by

\begin{equation}
\textnormal{\textbf{W}}=\text{\ensuremath{\mathbb{E}\left\{ \boldsymbol{\textnormal{\textbf{H}}}_{k}^{\textnormal{llc}}\boldsymbol{\textnormal{\textbf{v}}}_{k}^{\textnormal{llc}}\left(\,\boldsymbol{\textnormal{\textbf{H}}}_{k}^{\textnormal{llc}}\boldsymbol{\textnormal{\textbf{v}}}_{k}^{\textnormal{llc}}\right)^{\textnormal{H}}\right\} }}+\sigma^{^{2}}\boldsymbol{\textnormal{I}}_{M_{r}},
\end{equation}
where $\boldsymbol{\textnormal{I}}_{M_{r}}$ is $M_{r}\times M_{r}$
identity matrix. The IRC vector $\left(\boldsymbol{\textnormal{\textbf{u}}}_{k}^{\textnormal{llc}}\right)^{(1)}$
is then de-oriented to be aligned within one possible null space of
the effective inter-user interference subspace $\boldsymbol{\textnormal{\textbf{H}}}_{k}^{\textnormal{llc}}\boldsymbol{\textnormal{\textbf{v}}}_{\textnormal{ref}}$,
as expressed by

\begin{equation}
\left(\boldsymbol{\textnormal{\textbf{u}}}_{k}^{\textnormal{llc}}\right)^{(2)}=\left(\boldsymbol{\textnormal{\textbf{u}}}_{k}^{\textnormal{llc}}\right)^{(1)}-\frac{\left(\left(\boldsymbol{\textnormal{\textbf{u}}}_{k}^{\textnormal{llc}}\right)^{(1)}\cdotp\,\boldsymbol{\,\,\textnormal{\textbf{H}}}_{k}^{\textnormal{llc}}\boldsymbol{\textnormal{\textbf{v}}}_{\textnormal{ref}}\right)}{\left\Vert \boldsymbol{\textnormal{\textbf{H}}}_{k}^{\textnormal{llc}}\boldsymbol{\textnormal{\textbf{v}}}_{\textnormal{ref}}\right\Vert ^{2}}\times\boldsymbol{\textnormal{\textbf{H}}}_{k}^{\textnormal{llc}}\boldsymbol{\textnormal{\textbf{v}}}_{\textnormal{ref}}.
\end{equation}

This way, the final URLLC decoder vector $\left(\boldsymbol{\textnormal{\textbf{u}}}_{k}^{\textnormal{llc}}\right)^{(2)}$
exhibits no inter-user interference, providing the URLLC user with
a robust decoding ability. 

\subsection{Analytic Analysis Compared to State of The Art }

We compare the performance of the proposed NSBPS scheduler against
the state-of-the-art schedulers as follows:

\textbf{1. Punctured scheduler (PS)} {[}8{]}: the URLLC traffic is
always assigned a higher scheduling priority. If radio resources are
not available, PS scheduler instantly overwrites part of the ongoing
eMBB transmissions, i.e., immediately stop an ongoing eMBB transmission,
for instant URLLC scheduling. PS scheduler shows significant improvement
of the URLLC latency performance at the expense of highly degraded
SE.

\textbf{2. Multi-user punctured scheduler (MUPS)} {[}14{]}: in our
past work, we considered a MU scheduler on top of the PS scheduler.
MUPS first attempts to achieve a successful MU-MIMO transmission between
a URLLC-eMBB user pair; however, it is a transparent, non-biased and
non-controlled MU-MIMO. If the SDoFs are limited, MUPS scheduler rolls
back to PS scheduler. MUPS has shown an improved performance tradeoff
between system SE and URLLC latency; however, with a limited and non-robust
gain, due to the non-controlled MU-MIMO and the SE-less efficient
PS events. 

Accordingly, the aggregate eMBB user rate can be linearly calculated
from the individual sub-carrier rates for simplicity, assuming OFDMA
flat fading channels, as 

\begin{equation}
r_{_{k}}^{\textnormal{mbb}}=\Xi_{k}^{\textnormal{mbb}}r_{_{k,rb}}^{\textnormal{mbb}}.
\end{equation}

Then, the portion of the radio resources $\varGamma_{k_{\textnormal{mbb}}}^{\textnormal{llc}}$
allocated to the $k^{th}$ eMBB user, and being altered by the sporadic
URLLC traffic, can be expressed by a set of random variables, as

\begin{equation}
\boldsymbol{\Gamma}=\left(\varGamma_{k_{\textnormal{mbb}}}^{\textnormal{llc}}\mid k_{\textnormal{mbb}}\in\text{\ensuremath{\mathcal{K}}}_{\textnormal{mbb}}\right).
\end{equation}
 Since URLLC packets are of small payload size, it is reasonably to
assume that $\varGamma_{k_{\textnormal{mbb}}}^{\textnormal{llc}}\leq\Xi_{k}^{\textnormal{mbb}}$
is almost surely satisfied. Hence, the actual eMBB rate is formulated
by the joint URLLC-eMBB rate allocation function, given by 

\begin{equation}
R_{k_{\textnormal{mbb}}}=\mathcal{F}\left(\Xi_{k}^{\textnormal{mbb}},\varGamma_{k_{\textnormal{mbb}}}^{\textnormal{llc}}\right).
\end{equation}

For an instance, if an eMBB user is allocated SU dedicated resources,
then $\mathcal{F}\left(\Xi_{k}^{\textnormal{mbb}},\varGamma_{k_{\textnormal{mbb}}}^{\textnormal{llc}}\right)=\Xi_{k}^{\textnormal{mbb}}\,r_{_{k,rb}}^{\textnormal{mbb}}$
with no capacity loss. However, due to the prioritized URLLC traffic,
the actual eMBB user rate suffers a loss over a portion of the allocated
resources, expressed by the rate loss function $\text{\ensuremath{\mathit{\textnormal{\ensuremath{\Pi}}}}}$
as

\begin{equation}
\mathcal{F}\left(\Xi_{k}^{\textnormal{mbb}},\varGamma_{k_{\textnormal{mbb}}}^{\textnormal{llc}}\right)=\Xi_{k}^{\textnormal{mbb}}r_{_{k,rb}}^{\textnormal{mbb}}\left(1-\textnormal{\ensuremath{\Pi}}\right),
\end{equation}
where the rate loss function $\textnormal{\ensuremath{\Pi}}:\left[0,1\right]\rightarrow\left[0,1\right]$
indicates the effective portion of impacted PRBs of the $k^{th}$
eMBB user. Under the proposed NSBPS scheduler, the gain of the updated
eMBB effective channel is given as

\begin{equation}
\text{\ensuremath{\mathcal{Q}}}_{k}^{\textnormal{mbb}}=\frac{1}{\left[\left(\boldsymbol{\textnormal{\textbf{H}}}_{k}^{\textnormal{mbb}}\left(\boldsymbol{\textnormal{\textbf{v}}}_{k^{\diamond}}^{\textnormal{mbb}}\right)^{\prime}\right)\times\left(\boldsymbol{\textnormal{\textbf{H}}}_{k}^{\textnormal{mbb}}\left(\boldsymbol{\textnormal{\textbf{v}}}_{k^{\diamond}}^{\textnormal{mbb}}\right)^{\prime}\right)^{\textnormal{H}}\right]^{-1}},
\end{equation}
where $\text{\ensuremath{\mathcal{Q}}}_{k}^{\textnormal{mbb}}$ is
the achievable post-projection channel gain of the $k^{th}$ eMBB
user, and its magnitude can be rewritten in terms of the precoder
projection loss, i.e., the \textit{on-the-fly} eMBB precoder update
from $\boldsymbol{\textnormal{\textbf{v}}}_{k^{\diamond}}^{\textnormal{mbb}}$
to $\left(\boldsymbol{\textnormal{\textbf{v}}}_{k^{\diamond}}^{\textnormal{mbb}}\right)^{\prime},$
as

\begin{equation}
\text{\ensuremath{\mathcal{Q}}}_{k}^{\textnormal{mbb}}=\left\Vert \boldsymbol{\textnormal{\textbf{H}}}_{k}^{\textnormal{mbb}}\boldsymbol{\textnormal{\textbf{v}}}_{k^{\diamond}}^{\textnormal{mbb}}\right\Vert ^{2}\times\sin^{2}\left(\theta_{\left[\boldsymbol{\textnormal{\textbf{v}}}_{k^{\diamond}}^{\textnormal{mbb}},\left(\boldsymbol{\textnormal{\textbf{v}}}_{k^{\diamond}}^{\textnormal{mbb}}\right)^{\prime}\right]}\right),
\end{equation}
where $\sin^{2}\left(\theta_{\left[\boldsymbol{\textnormal{\textbf{v}}}_{k^{\diamond}}^{\textnormal{mbb}},\left(\boldsymbol{\textnormal{\textbf{v}}}_{k^{\diamond}}^{\textnormal{mbb}}\right)^{\prime}\right]}\right)$
introduces the eMBB projection loss, over the shared resources with
the URLLC traffic, with $\theta_{\left[\boldsymbol{\textnormal{\textbf{v}}}_{k^{\diamond}}^{\textnormal{mbb}},\left(\boldsymbol{\textnormal{\textbf{v}}}_{k^{\diamond}}^{\textnormal{mbb}}\right)^{\prime}\right]}$
as the spatial angle deviation between its original and projected
precoders. Thus, $\overset{\textnormal{NSBPS}}{\text{\ensuremath{\mathit{\textnormal{\ensuremath{\Pi}}}}}}$
can be expressed as

\begin{equation}
\overset{\textnormal{NSBPS}}{\text{\ensuremath{\mathit{\textnormal{\ensuremath{\Pi}}}}}}=\left(\frac{\varGamma_{k_{\textnormal{mbb}}}^{\textnormal{llc}}}{\Xi_{k}^{\textnormal{mbb}}}\right)\times\sin^{2}\left(\theta_{\left[\boldsymbol{\textnormal{\textbf{v}}}_{k^{\diamond}}^{\textnormal{mbb}},\left(\boldsymbol{\textnormal{\textbf{v}}}_{k^{\diamond}}^{\textnormal{mbb}}\right)^{\prime}\right]}\right).
\end{equation}

Due to the constraints in (14) and (18), the projection loss is always
 guaranteed minimum, i.e., $\sin^{2}\left(\theta_{\left[\boldsymbol{\textnormal{\textbf{v}}}_{k^{\diamond}}^{\textnormal{mbb}},\left(\boldsymbol{\textnormal{\textbf{v}}}_{k^{\diamond}}^{\textnormal{mbb}}\right)^{\prime}\right]}\right)\ll1$.
For the PS scheduler, the rate loss function is expressed in terms
of the entire URLLC resources inducing the resource allocation of
the eMBB user, since the eMBB transmission is instantly stopped over
these resources, as

\begin{equation}
\overset{\textnormal{PS}}{\textnormal{\ensuremath{\Pi}}}=\left(\frac{\varGamma_{k_{\textnormal{mbb}}}^{\textnormal{llc}}}{\Xi_{k}^{\textnormal{mbb}}}\right).
\end{equation}

Finally, the MUPS scheduler exhibits an average eMBB capacity loss
due to the persistent PS events, if the normal MU-MIMO scheduler fails;
thus, the rate loss can be given as

\begin{equation}
\overset{\textnormal{MUPS}}{\textnormal{\ensuremath{\Pi}}}=\varPhi\left(\frac{\varGamma_{k_{\textnormal{mbb}}}^{\textnormal{llc}}}{\Xi_{k}^{\textnormal{mbb}}}\right),
\end{equation}
where $\varPhi\leq1$ is a fraction to indicate the probability density
of rolling back to PS scheduler, under a specific cell loading. Hence,
the average eMBB user rate can be calculated as

\begin{equation}
\overline{R}_{k_{\textnormal{mbb}}}=\Xi_{k}^{\textnormal{mbb}}r_{_{k,rb}}^{\textnormal{mbb}}\left(1-\text{\ensuremath{\mathbb{E}\left\{ \Pi\right\} }}\right).
\end{equation}

Based on (26) - (33), it can be concluded that the proposed NSBPS
scheduler provides the best achievable eMBB and URLLC joint performance
against state-of-the-art schedulers. 

\section{Simulation Results}

In this section, we present the extensive SLS results of the NSBPS
scheduler, following the 5G-NR specifications, 
\begin{table}
\caption{Simulation Parameters.}
\centering{}%
\begin{tabular}{c|c}
\hline 
Parameter & Value\tabularnewline
\hline 
Environment & $\begin{array}{c}
\textnormal{3GPP-UMA,7 gNBs, 21 cells,}\\
\textnormal{500 meters inter-site distance}
\end{array}$\tabularnewline
\hline 
Channel bandwidth & 10 MHz, FDD\tabularnewline
\hline 
Antenna setup & BS: 8 Tx, UE: 2 Rx\tabularnewline
\hline 
User dropping & $\begin{array}{c}
\textnormal{uniformly distributed}\\
\textnormal{URLLC: 5, 10 and 20 users/cell}\\
\textnormal{eMBB: 5 , 10 and 20 users/cell}
\end{array}$\tabularnewline
\hline 
User receiver & LMMSE-IRC\tabularnewline
\hline 
TTI configuration & $\begin{array}{c}
\textnormal{URLLC: 0.143 ms (2 OFDM symbols)}\\
\textnormal{eMBB: 1 ms (14 OFDM symbols)}
\end{array}$\tabularnewline
\hline 
CQI & periodicity: 5 ms, with 2 ms latency\tabularnewline
\hline 
HARQ & $\begin{array}{c}
\textnormal{asynchronous HARQ, Chase combining}\\
\textnormal{HARQ round trip time = 4 TTIs}
\end{array}$\tabularnewline
\hline 
Link adaptation & $\begin{array}{c}
\textnormal{dynamic modulation and coding}\\
\textnormal{target URLLC BLER : 1\%}\\
\textnormal{target eMBB BLER : 10\%}
\end{array}$\tabularnewline
\hline 
Traffic model & $\begin{array}{c}
\textnormal{\textnormal{URLLC: bursty, }B=50 bytes, \ensuremath{\lambda=250} }\\
\textnormal{eMBB: full buffer}
\end{array}$\tabularnewline
\hline 
\end{tabular}
\end{table}
 where the main simulation parameters are listed in Table I. 

Fig. 2 shows the URLLC average one-way latency $\varPsi$ at the $10^{-5}$
outage probability, under proposed NSBPS, PS, MUPS, and WPF schedulers.
On the top left, a close snap of the complementary cumulative distribution
function (CCDF) of the URLLC latency distribution is further presented.
We define the cell load setup by: $\varOmega=(K_{\textnormal{mbb}},\,K_{\textnormal{llc}}).$
The proposed NSBPS scheduler clearly provides a significantly robust
and steady URLLC latency against different cell load conditions, and
hence, independently from the aggregate levels of interference. The
overall performance gain of the NSBPS scheduler is due to: 1) the
guaranteed instantaneous URLLC scheduling without queuing in a controlled
(almost surely occurs), biased (for the sake of the URLLC user), and
semi-transparent (URLLC user is aware of it) MU transmission, leading
to no inter-user interference at the URLLC user, 2) the constrained-minimum
eMBB user rate loss function, and 3) the enforced regularization of
the inter-cell interference spatial distribution within a limited
span, due to the fixed subspace projection, and hence, the linear
MMSE-IRC receiver nulls the average inter-cell interference more efficiently
and with 
\begin{figure}
\begin{centering}
\includegraphics[scale=0.49]{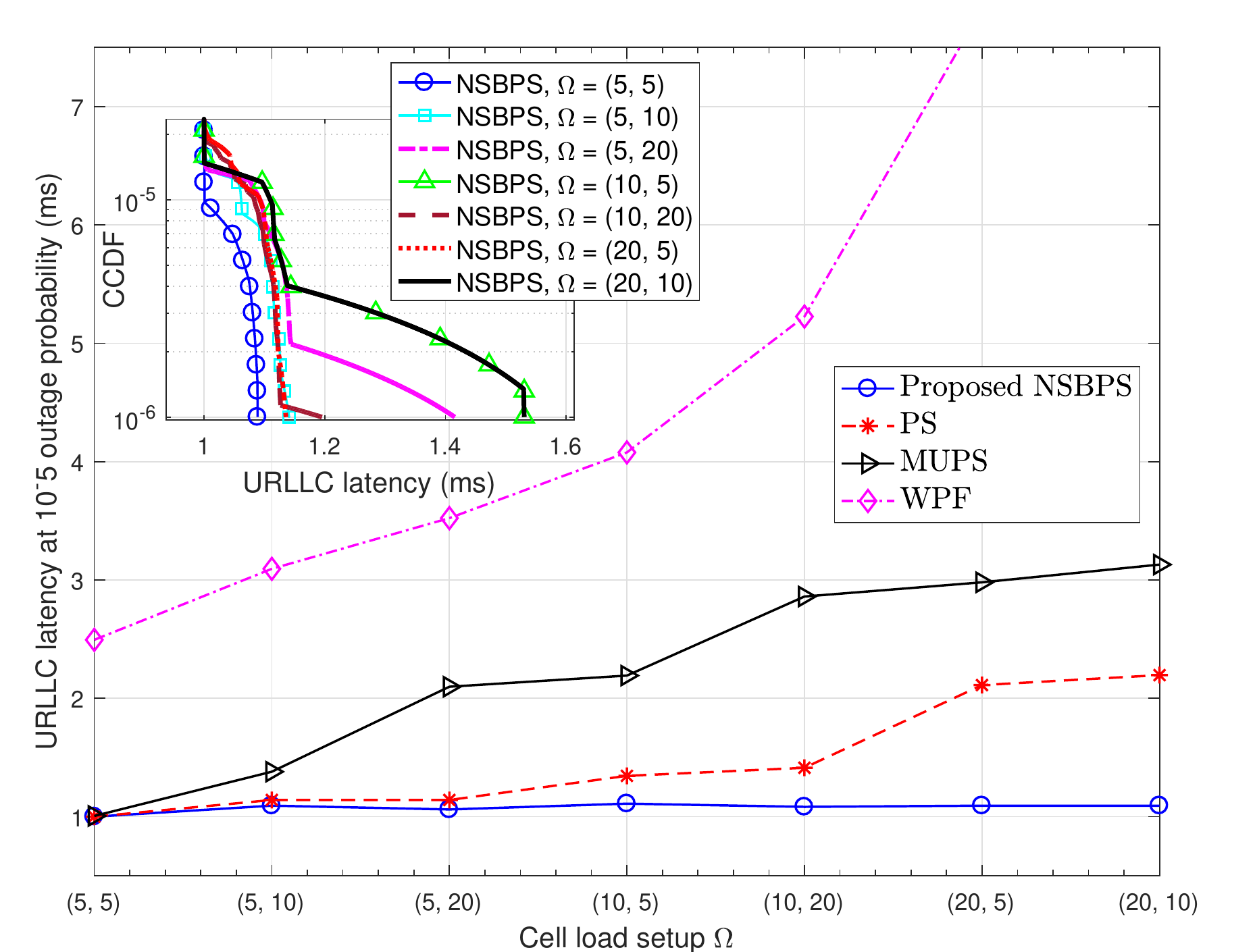}
\par\end{centering}
\centering{}{\small{}~~~~~~~Fig. 2. URLLC one-way latency }$\varPsi${\small{}
at $10^{-5}$outage.}{\small \par}
\end{figure}
 improved SDoFs. 

The PS scheduler shows an optimized URLLC latency in the low load
region, at the expense of degraded eMBB performance. However, in the
high load region when the inter-cell interference levels are extreme,
PS scheduler provides a degraded URLLC latency performance due to
the experienced re-transmissions and degraded capacity per PRB. The
MUPS scheduler shows a fair tradeoff between URLLC latency and the
eMBB SE, where the non-controlled URLLC-eMBB MU-MIMO transmissions
reduce the URLLC decoding ability. Finally, the WPF scheduler exhibits
the worst URLLC latency performance, where the URLLC packets are queued
for multiple TTIs if the radio resources are not instantly schedulable. 

\begin{figure}
\begin{centering}
\includegraphics[scale=0.62]{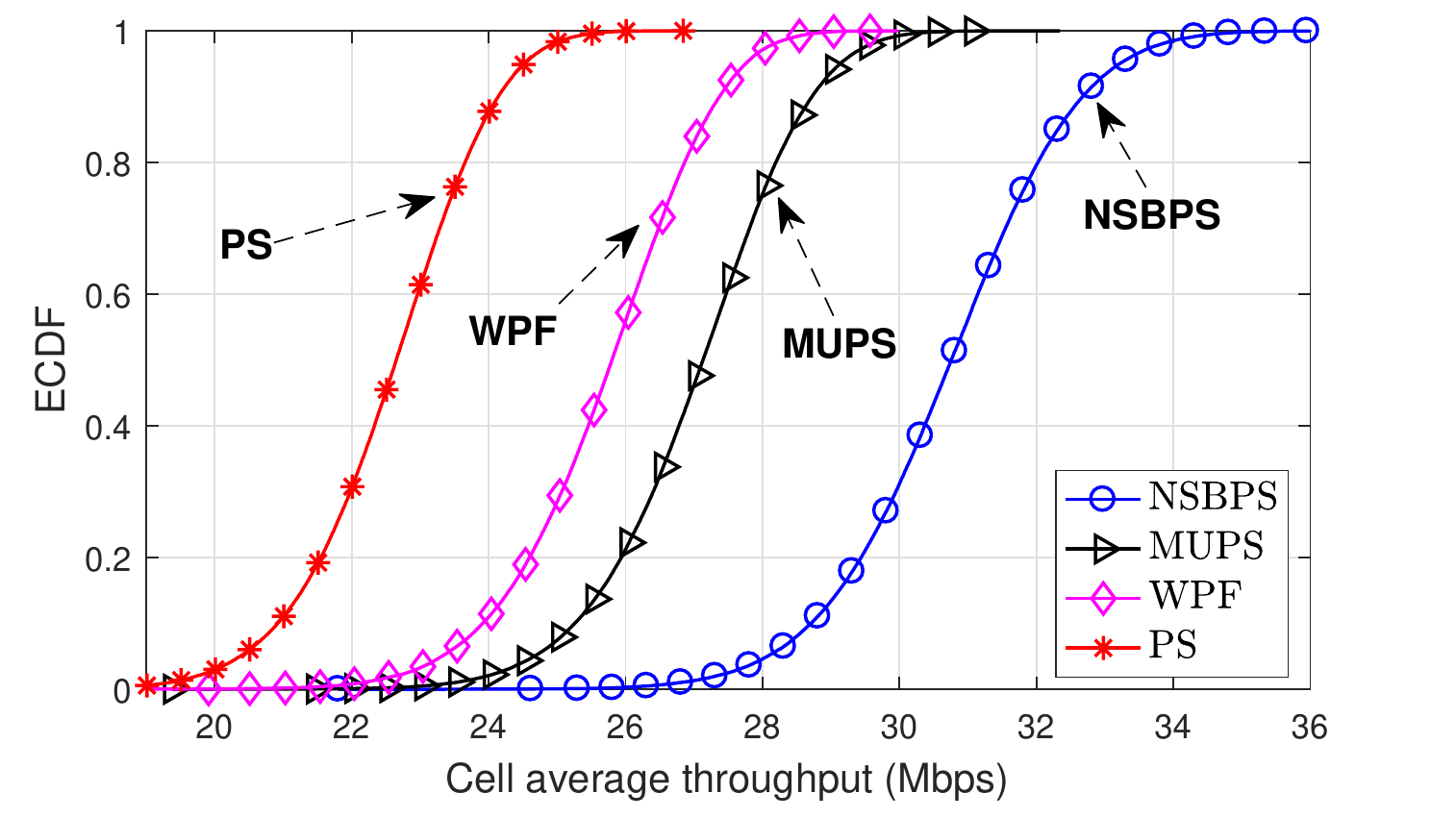}
\par\end{centering}
\centering{}{\small{}~~~~~~Fig. 3. Cell average throughput for
$\varOmega=(5,\,5).$}{\small \par}
\end{figure}
As shown in Fig. 3, the empirical CDF (ECDF) of the average cell throughput
in Mbps is presented. The NSBPS scheduler provides the best achievable
cell throughput compared to other schedulers, due to the always constrained-minimum
rate loss function of the victim eMBB users. The PS scheduler exhibits
severe loss of the network SE due to the punctured eMBB transmissions.
However, the WPF scheduler achieves an improved capacity since no
puncture-events are allowed; however, at the expense of the worst
URLLC latency. Finally, the MUPS scheduler shows further improved
capacity, due to the successful MU events; however, with limited MU
gain since when a successful MU pairing is not possible, MUPS falls
back to SE-less-efficient PS scheduler. 

Examining the eMBB performance, Fig. 4. shows the average eMBB user
throughput in Mbps, for all schedulers under evaluation, where similar
conclusions can be clearly obtained. For instance, with $\varOmega=(5,5)$,
where the system SDoFs are limited by the small number of active eMBB
users, i.e., $K_{\textnormal{mbb}}=5$, the proposed NSBPS shows a
gain $\sim28.9\%$ in the eMBB user throughput than the MUPS scheduler.
Under such SDoF-limited state, the MUPS scheduler is highly likely
to roll back to PS scheduler, i.e., $\varPhi\sim1,$ while the NSBPS
forcibly enforces these missing SDoFs, sufficient enough to instantly
fit the URLLC traffic within an eMBB transmission.

Finally, Table II presents the achievable MU throughput gain of the
NSBPS and MUPS schedulers. The best achievable MU gain of the NSBPS
over the MUPS scheduler is obtained when the system is originally
SDoF-limited, i.e., $\varOmega=(5,5)$. With SDoF-rich loading states
such as $\varOmega=(20,5)$, the MUPS scheduler rarely falls back
to PS scheduler, i.e., $\varPhi\sim0,$ and hence, an improved MU
gain is achieved. 

\section{Conclusion }

A null space based preemptive scheduler (NSBPS) has been proposed
for joint 5G URLLC and eMBB traffic. The proposed NSBPS scheduler
aims to fulfill a constraint-coupled objective, for which the URLLC
quality of service is almost surely guaranteed while achieving the
maximum possible ergodic capacity. Extensive system level simulations
and analytic gain analysis have been conducted for performance evaluation.
Compared to the state-of-the-art scheduler proposals from academia
and industry, the proposed NSBPS shows extreme robustness of the URLLC
latency performance, i.e., regardless of the cell loading, and aggregate
interference levels, while providing significantly improved eMBB performance.
A comprehensive study on the performance of the proposed 
\begin{figure}
\begin{centering}
\includegraphics[scale=0.62]{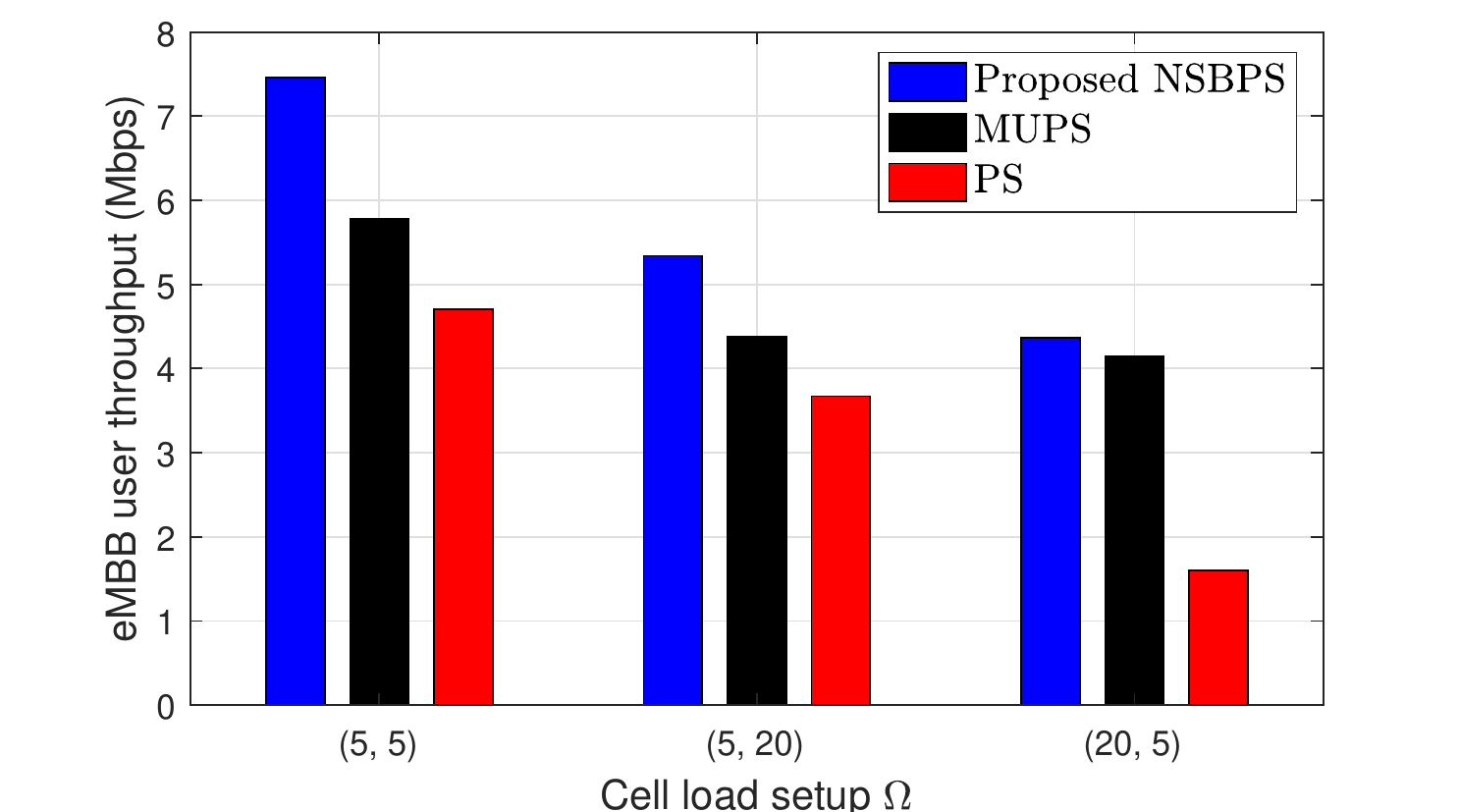}
\par\end{centering}
\centering{}{\small{}~~~~~~Fig. 4. eMBB average user throughput.}{\small \par}
\end{figure}
\begin{table}
\caption{Average MU gain of the NSBPS and MUPS schedulers.}
\centering{}%
\begin{tabular}{c|c|c|c}
\hline 
Scheduler & $\varOmega=(5,5)$ & $\varOmega=(5,20)$ & $\varOmega=(20,5)$\tabularnewline
\hline 
MUPS, MU gain (Mbps)  & 7.69 & 12.13 & 23.05\tabularnewline
\hline 
NSBPS, MU gain (Mbps)  & 22.92 & 24.91 & 27.78\tabularnewline
\hline 
Gain (\%) & +198.04 & +105.35 & +20.52\tabularnewline
\hline 
\end{tabular}
\end{table}
 scheduler will be considered in a future work.


\begin{thebibliography}{10}
\bibitem[1]{key-1}NR and NG-RAN overall description; Stage-2 (Release
15), 3GPP, TS 38.300, V2.0.0, Dec. 2017.

\bibitem[2]{key-2}Service requirements for the 5G system; Stage-1
(Release 16), 3GPP, TS 22.261, V16.2.0, Dec. 2017.

\bibitem[3]{key-3}Study on new radio access technology; Radio access
architecture and interfaces (Release 14), 3GPP, TR 38.801, V14.0.0,
March 2017.

\bibitem[4]{key-4}B. Soret, P. Mogensen, K. I. Pedersen and M. Aguayo-Torres,
\textquotedbl{}Fundamental tradeoffs among reliability, latency and
throughput in cellular networks,\textquotedbl{} \textit{in Proc. IEEE
Globecom}, Austin, TX, 2014, pp. 1391-1396.

\bibitem[5]{key-5}K. I. Pedersen, G. Berardinelli, F. Frederiksen,
P. Mogensen and A. Szufarska, \textquotedbl{}A flexible 5G frame structure
design for FDD cases,\textquotedbl{} \textit{IEEE Commun. Mag.}, vol.
54, no. 3, pp. 53-59, March 2016.

\bibitem[6]{key-6}G. Pocovi, B. Soret, M. Lauridsen, K. I. Pedersen
and P. Mogensen, \textquotedbl{}Signal quality outage analysis for
URLLC in cellular networks,\textquotedbl{} \textit{in Proc. IEEE Globecom},
San Diego, CA, 2015, pp. 1-6.

\bibitem[7]{key-7}J. J. Nielsen, R. Liu and P. Popovski, \textquotedbl{}Ultra-reliable
low latency communication using interface diversity,\textquotedbl{}
\textit{IEEE Trans. Commun,} vol. 66, no. 3, pp. 1322-1334, March
2018. 

\bibitem[8]{key-8}K.I. Pedersen, G. Pocovi, J. Steiner, and S. Khosravirad,
\textquotedblleft Punctured scheduling for critical low latency data
on a shared channel with mobile broadband,\textquotedblright{} \textit{in
Proc. IEEE VTC}, Toronto, 2017, pp. 1-6.

\bibitem[9]{key-9}Study on 3D channel model for LTE; Release 12,
3GPP, TR 36.873, V12.7.0, Dec. 2014

\bibitem[10]{key-10}Y. Ohwatari, N. Miki, Y. Sagae and Y. Okumura,
\textquotedbl{}Investigation on interference rejection combining receiver
for space\textendash frequency block code transmit diversity in LTE-advanced
downlink,\textquotedbl{} \textit{IEEE Trans. Veh. Technol.}, vol.
63, no. 1, pp. 191-203, Jan. 2014

\bibitem[11]{key-11}S. N. Donthi and N. B. Mehta, \textquotedbl{}An
accurate model for EESM and its application to analysis of CQI feedback
schemes,\textquotedbl{} \textit{IEEE Trans. Wireless Commun.}, vol.
10, no. 10, pp. 3436-3448, Oct. 2011. 

\bibitem[12]{key-12}Physical layer procedures; Evolved universal
terrestrial radio access (Release 15), 3GPP, TS 36.213, V15.1.0, March.
2018.

\bibitem[13]{key-13}Bertsekas, D. and Gallager, R. (1992). \textit{Data
Networks}. 2nd ed. Michigan: Prentice Hall.

\bibitem[14]{key-14}Ali A. Esswie, and K.I. Pedersen, \textquotedblleft Multi-user
preemptive scheduling for critical low latency communications in 5G
networks,\textquotedblright{} \textit{in Proc. IEEE ISCC}, Natal,
2018, pp. 1-6.
\end{thebibliography}
\end{document}